\pdfoutput=1
\documentclass[conference,a4paper]{IEEEtran}

\usepackage{subcaption}
\usepackage{amsmath}
\ifCLASSINFOpdf
   \usepackage[pdftex]{graphicx}
\else
\fi
\begin{document}
%
\title{Integration of Renewable Power Sources
into the Vietnamese Power System }

\author{
    \IEEEauthorblockN{Alexander Kies\IEEEauthorrefmark{1}, Bruno Schyska\IEEEauthorrefmark{2}, Dinh Thanh Viet\IEEEauthorrefmark{3}, Lueder von Bremen\IEEEauthorrefmark{2},
    Detlev Heinemann\IEEEauthorrefmark{2},Stefan Schramm\IEEEauthorrefmark{1}}
    \IEEEauthorblockA{\IEEEauthorrefmark{1} Frankfurt Institute for Advanced Studies, Goethe University Frankfurt, Ruth-Moufang-Str. 1, 60438 Frankfurt am Main, Germany
    }
    \IEEEauthorblockA{\IEEEauthorrefmark{2} ForWind, Center for Wind Energy Research, University of Oldenburg, Kuepkersweg 70, 26129 Oldenburg, Germany
    }
    \IEEEauthorblockA{\IEEEauthorrefmark{3} Department of Electrical Engineering, University of Danang, 41 Le Duan St, Danang, Vietnam
    }
}


%


\maketitle

\begin{abstract}
The Vietnamese Power system is expected to expand considerably in upcoming decades.
However, pathways towards higher shares of renewables ought to be investigated.
In this work, we investigate a highly renewable Vietnamese power system by jointly optimising the expansion of renewable
generation facilities and the transmission grid.
We show that in the cost-optimal case, highest amounts of wind capacities are installed in southern Vietnam and solar photovoltaics (PV) in central Vietnam.
In addition, we show that transmission has the potential to reduce levelised cost of electricity by approximately 10\%.
\end{abstract}


\section{Introduction}
Power systems are transforming worldwide.
The transformation from conventional dispatchable power generation from fossil sources towards power generation from the renewable sources of mainly wind, solar and hydro is driven by 
goals of sustainability and reduction of climate gas emissions to mitigate climate change.
However, renewable power generation depends on the weather and therefore has strongly fluctuating feed-in profiles which, in turn, make the system integration of renewables difficult.
Among solutions to integrate high shares of renewable into power systems are:
i) optimising the mix of generation from different renewable sources  (\cite{Lund_2006,franccois2016increasing,Kies_2015,santos2015combining,kies2016optimal,jurasz2017modeling})
ii) storage (\cite{heide2011reduced, chattopadhyay2017impact})
iii) dispatchable backup power (\cite{schlachtberger2016backup, plessmann2014global})
iv) sector coupling (\cite{brown2017sector, schaber2013managing})
v) transmission grid extensions (\cite{brown2014transmission, becker2014transmission})
vi) controllable hydro power (\cite{kies2016effect})
vii) system-friendly renewables (\cite{hirth2016system, kies2016curtailment})
or 
viii) demand-side management (\cite{palensky2011demand, kies2016demand}).
\\
In developing countries such as Vietnam, power demand grows strongly and a reliable energy supply has imminent importance for
stable economic growth and prosperity. The Vietnamese demand for electricity is growing at an average speed of 7-8\%
per year and the peak demand is estimated to reach 42.1 GW by 2020 and 90.7 GW by 2030 (\cite{asianbank}). 
\\At present, Vietnam has exploited mainly thermal and hydro power resources. Coal as major energy carrier is being
imported from abroad. However, the number of thermal power plants is likely to be limited in the future due to concerns raised by environmental pollution and dependency on imports. Therefore, the importance of renewable energy
sources such as wind, solar, tidal, biomass will play an increasingly important role for Vietnam.
\\ The national energy development vision until 2050, which was approved by the Vietnamese Prime Minister, emphasises the role of renewable energy sources in particular. It is expected that by 2050, 43\% of Vietnam's electricity will
be provided from renewable sources. Expectations about installed wind power capacity are 800 MW in 2020, 2,000
MW in 2025 and around 6,000 MW by 2030. For solar energy, predicted numbers of installed capacities are 850 MW by 2020, 4,000
MW by 2025 and 12,000 MW by 2030. In addition, biomass will contribute about 1\% of entire generation by 2020,
1.2\% by 2025 and 2.1\% by 2030 (\cite{VietnamPDP}). \\
In 2014, the Vietnamese electricity production of 145.5 TWh was mainly supplied by hydro power (38\%, not including small hydro), CCGT (31\%) and coal (26\%). The dominance of those three sources was also reflected by their
shares in overall installed capacities of 34 GW, comprising hydro (40\%, not including small hydro), gas (22\%) and
coal (29\%). However, the Vietnamese power development plan predicts total installed capacities by 2030 of 116 GW with
shrinking shares of hydro (18\%) and gas (17\%) and growing shares of coal (50\%) and renewables (10\%) (\cite{Vietnam3}).
\\In this paper, we investigate the optimal mix of renewables from wind and photovoltaics distributed among the highest
voltage substations of a simpliﬁed future Vietnamese power system. \\
Similar studies of the power system transition of developing countries have been performed for a variety of countries such as Iran (\cite{aghahosseinirole}) and
also Vietnam (\cite{nguyen2005long}), but in the latter case with a strong focus on conventional generation technologies and without including time-dependent renewable resource availabilities. 

\section{ Methodology}
\begin{figure}[t]\vspace*{4pt}
\centerline{\includegraphics[width=.5\textwidth]{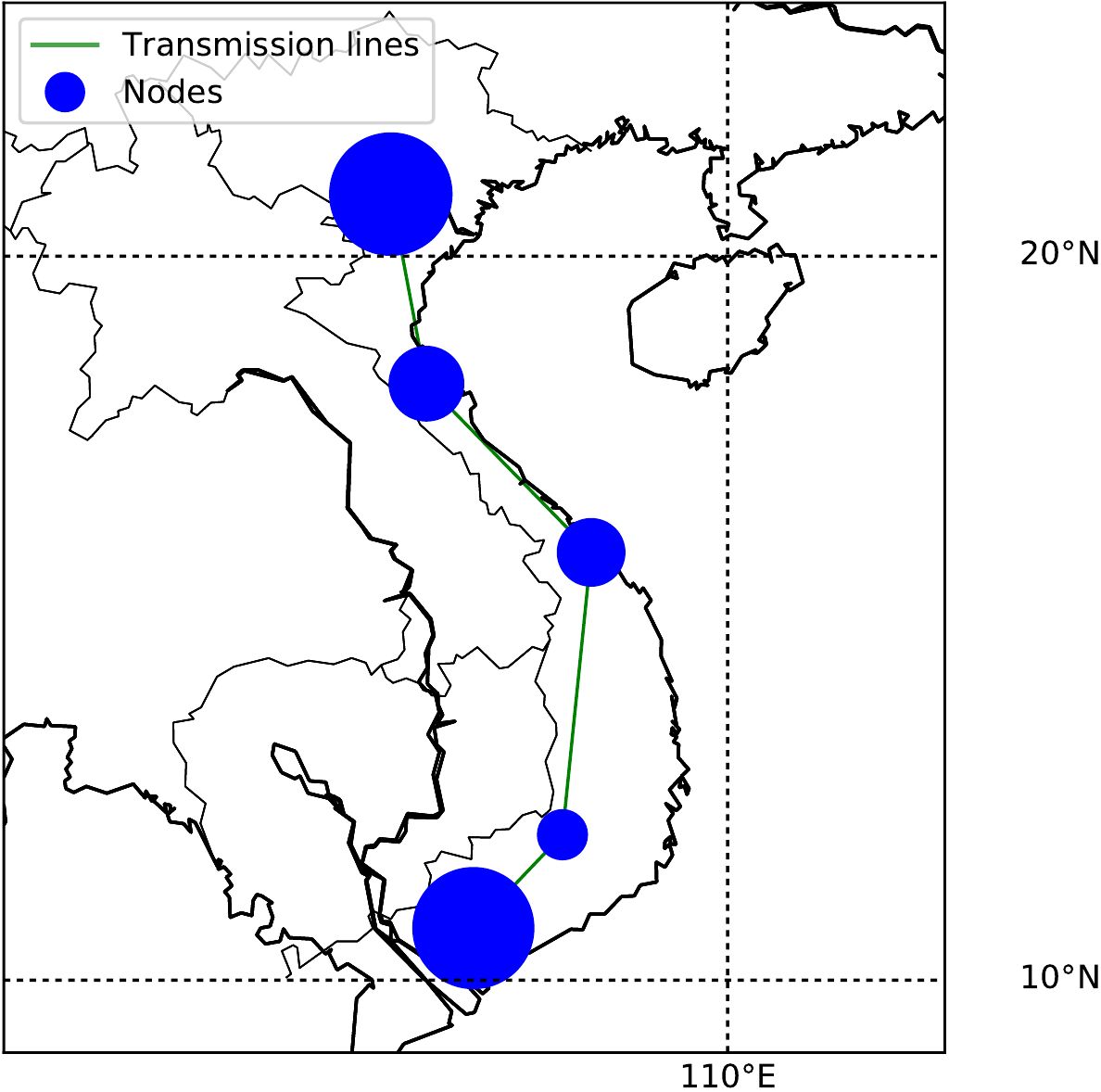}}
\caption{\label{fig:topology}Topology of the investigated simplified Vietnamese power system. The sizes of the circles indicate the relative average demand per node.}
\end{figure}

We use a simplified version of the Vietnamese power system, where loads and generation are connected to the closest existing highest voltage substation.
The topology of the resulting network is shown in Fig. \ref{fig:topology}. The model is formulated as a linear optimisation model that minimises total system cost. 
The objective reads 
\begin{equation} \label{eq:main}
\min_{g,G,f,F} (\sum_{n,s} c_{n,s} G_{n,s} + \sum_l c_l F_l + \sum_{n,s,t} o_{n,s} g_{n,s}(t)),
\end{equation}
where $c_{n,s}$ is the equivalent investment cost for generation capacity, $c_l$ is the equivalent investment cost for transmission capacity, $o_{n,s}$ is the marginal cost of energy generation,
$G_{n,s}$ and $F_l$ are the capacities of generators and transmission links and $g_{n,s}$ is the dispatch time series.
The index $n$ runs over all nodes and $s$ over considered technologies (wind, solar PV and OCGT).
\\In addition to the objective, multiple constraints have to be satisfied.
To ensure stable power system operation,
generation and demand need to match in space and time:
        \begin{equation}
\sum_s g_{n,s}(t) - d_n(t) = \sum_l K_{n,l} f_l(t) \forall n,t,
\end{equation}
where $d_{n}(t)$ is the demand, $K$ is the incidence matrix of the network and $f_l$ the flow over link $l$.\\
The dispatch of a generator $g_{n,s}(t)$ is constrained by the corresponding generator capacity $G_{n,s}$ multiplied with the corresponding hourly capacity factor $\bar{g}_{n,s}(t)$:
    \begin{equation}
    0 \leq g_{n,s}(t) \leq \bar{g}_{n,s}(t) G_{n,s}\forall n,s,t.
    \end{equation}
Flows between nodes can not exceed transmission limits,  
    \begin{equation}
    |f_l(t)| \leq F_l \forall l,t,
    \end{equation}
where $F_l$ denotes the transmission limit of link $l$ (e.g., due to thermal limits).
However, line capacities can be expanded by the model.\\
Lastly, in some cases a global
$CO_2$ emission constraint is enforced, 
    \begin{equation}
    \sum_{n,s,t} \frac{1}{\eta_{n,s}}g_{n,s}(t) e_{n,s} \leq CAP_{CO_2} \label{eq:co2}, \\
    \end{equation}
        where $\eta$ and $e$ denote the technology specific efficiency and CO$_2$ emissions. This constraint is varied in the results section to investigate its influence
        on the optimal mix and system cost.
\\The methodology is described in more detail by Schlachtberger et al. (\cite{schlachtberger2017benefits}).
We use the software toolbox Python for Power System Analysis (PyPSA, pypsa.org)  to perform the simulations. \\
The optimisation problem (Eq. \ref{eq:main}) is solved for two transmission grid scenarios. First, with optimised transmission, i.e., 
without a global limit on $\sum_l F_l$ and with neglible capital cost for transmission grid capacity expansion.
Second, without transmission, i.e., $F_l = 0 \forall l$.
\subsection{ Data and assumptions}
The model simulates a full year with hourly resolution.
Data is based on the meteorological year 2012.
We use weather data from the MERRA reanalysis (\cite{rienecker2011merra}) to model spatially resolved generation from the renewable sources of wind and PV.
Raw weather data (wind speeds, irradiation and temperature) are converted to power as described by Kies et al. (\cite{restorereport2016}).
For wind power calculations, the power curve of a Vestas V90 at 90 m hub height is used and wind speeds are extrapolated to the desired hub height.
Capacities within the areas aggregated to nodes are distributed homogeneously among the single cells.
To obtain
irradiation on the tilted modules for PV power calculations, the Klucher model is applied (\cite{klucher1979evaluation}).\\
Cost assumptions for different technologies used to perform the cost optimisation are shown in Table \ref{tab:costsassumptions}.
\begin{center}
\begin{table*}
\begin{center}
\begin{tabular}{ cccccc } 
 \hline
 technology & investment cost & marginal cost & lifetime & efficiency & CO$_2$ emissions  \\ 
  &  [USD/kW] & [USD/MWh] & [a] &  & [tons / MWh$_{thermal}$]\\ 
 \hline
 wind & 1300 & 0.02 & 20 & & \\ 
 solar & 660 & 0.03 & 20 & & \\ 
 OCGT & 440 & 64.0 & 30 & 0.39 & 0.1872 \\ 
 \hline
\end{tabular}
\caption{Cost assumptions for generation technologies based on 2030 value estimates from Schroeder et al. (\cite{schroeder2013current}). For transmission, plain cost of 0.01 USD/MVA was assumed for each link to ensure uniqueness of the solution.}
\label{tab:costsassumptions}
\end{center}
\end{table*} 
\end{center}
\section{ Results}
\begin{figure}[!htb]\vspace*{4pt}
\begin{subfigure}{0.5\textwidth}
\includegraphics[width=\textwidth]{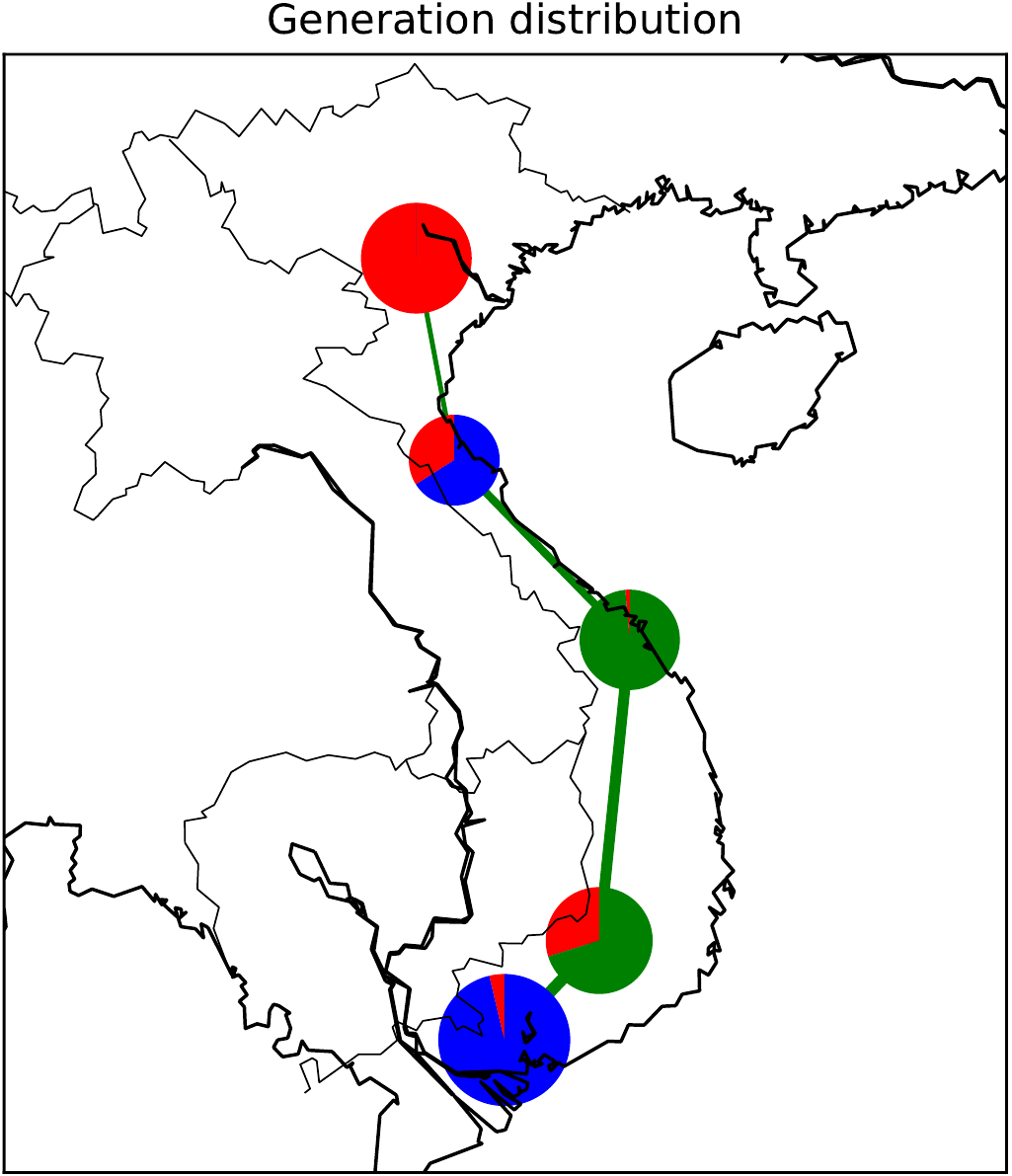}
\subcaption{With optimised transmission}
\end{subfigure}
\begin{subfigure}{0.5\textwidth}
\includegraphics[width=\textwidth]{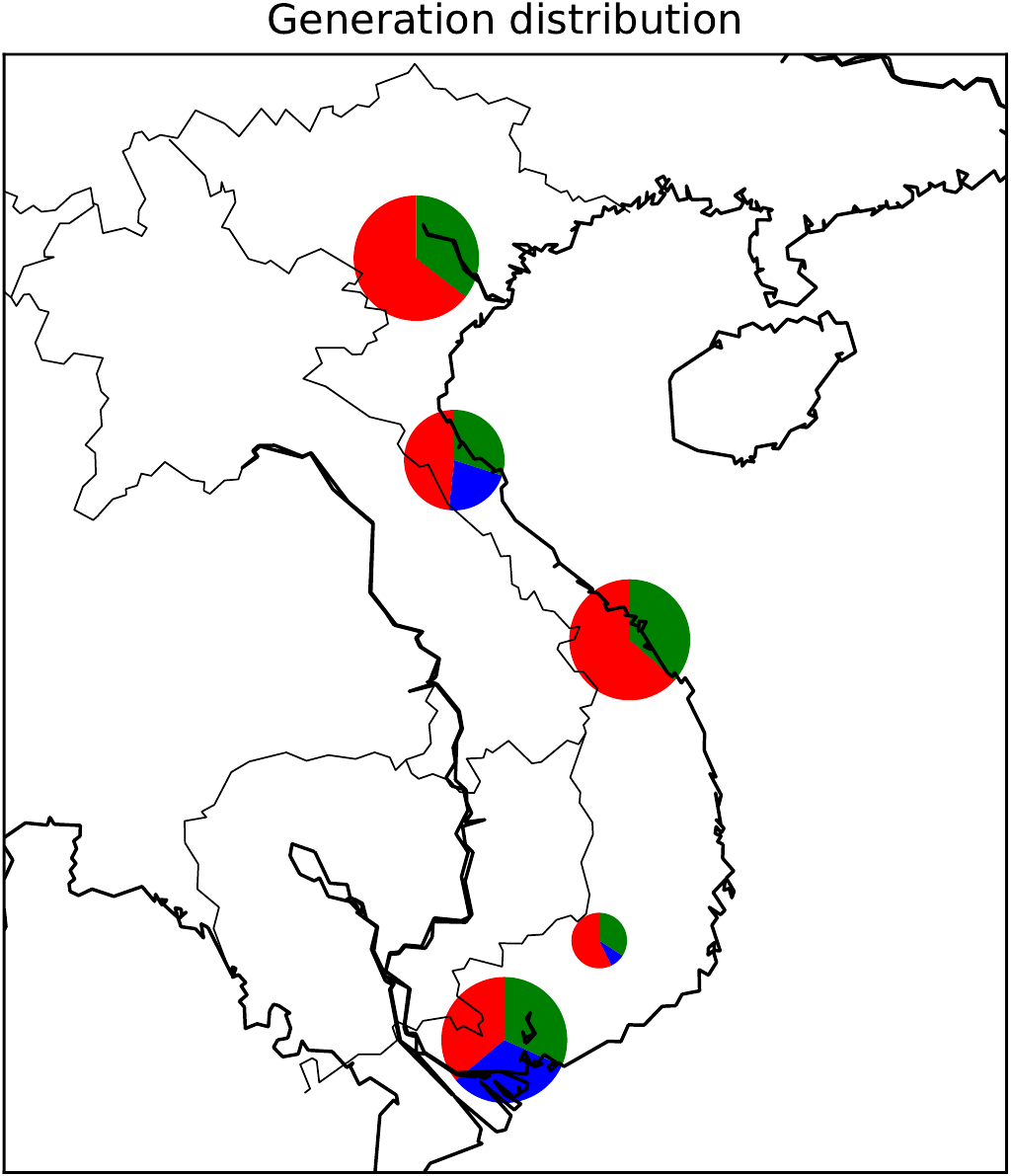}
\subcaption{Without transmission}
\end{subfigure}
\caption{\label{fig:distrb} Cost-optimal distribution and mix of generation. The colors indicate the shares of OCGT (red), wind (blue) and solar PV (green) and the sizes of the dots indicate overall generation.
Widths of transmission links indicate capacity of transmission link.}
\end{figure}

\begin{figure}[!htb]\vspace*{4pt}
\begin{subfigure}{0.5\textwidth}
\includegraphics[width=\textwidth]{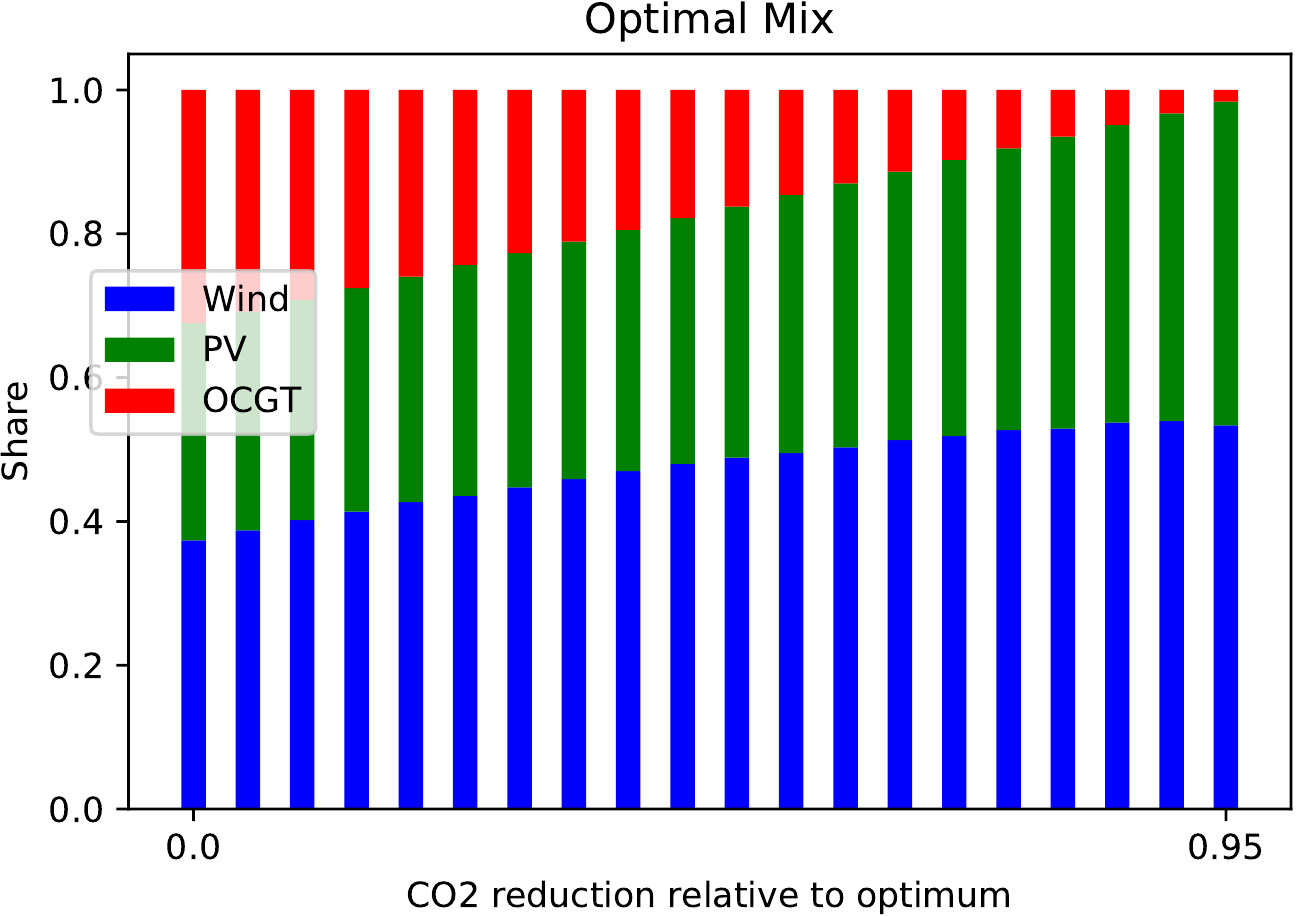}
\subcaption{With optimised transmission}
\end{subfigure}
\begin{subfigure}{0.5\textwidth}
 \includegraphics[width=\textwidth]{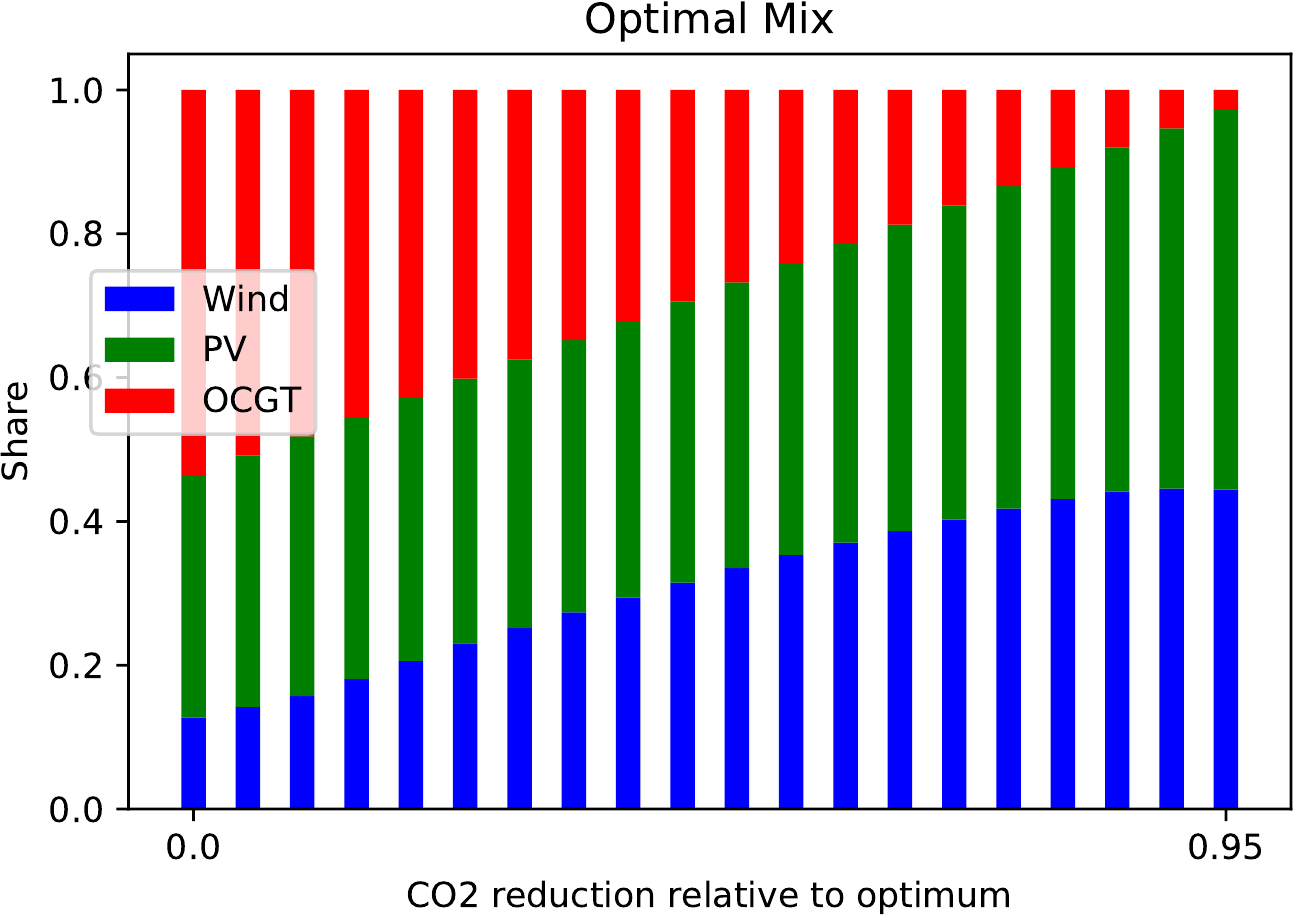}
 \subcaption{Without transmission}
\end{subfigure}
\caption{\label{fig:optmix} Optimal generation mix in dependency of the $CO_2$ reduction relative to the cost-optimum without any constraints.}
\end{figure}

\begin{figure}[!htb]\vspace*{4pt}
\centerline{\includegraphics[width=.5\textwidth]{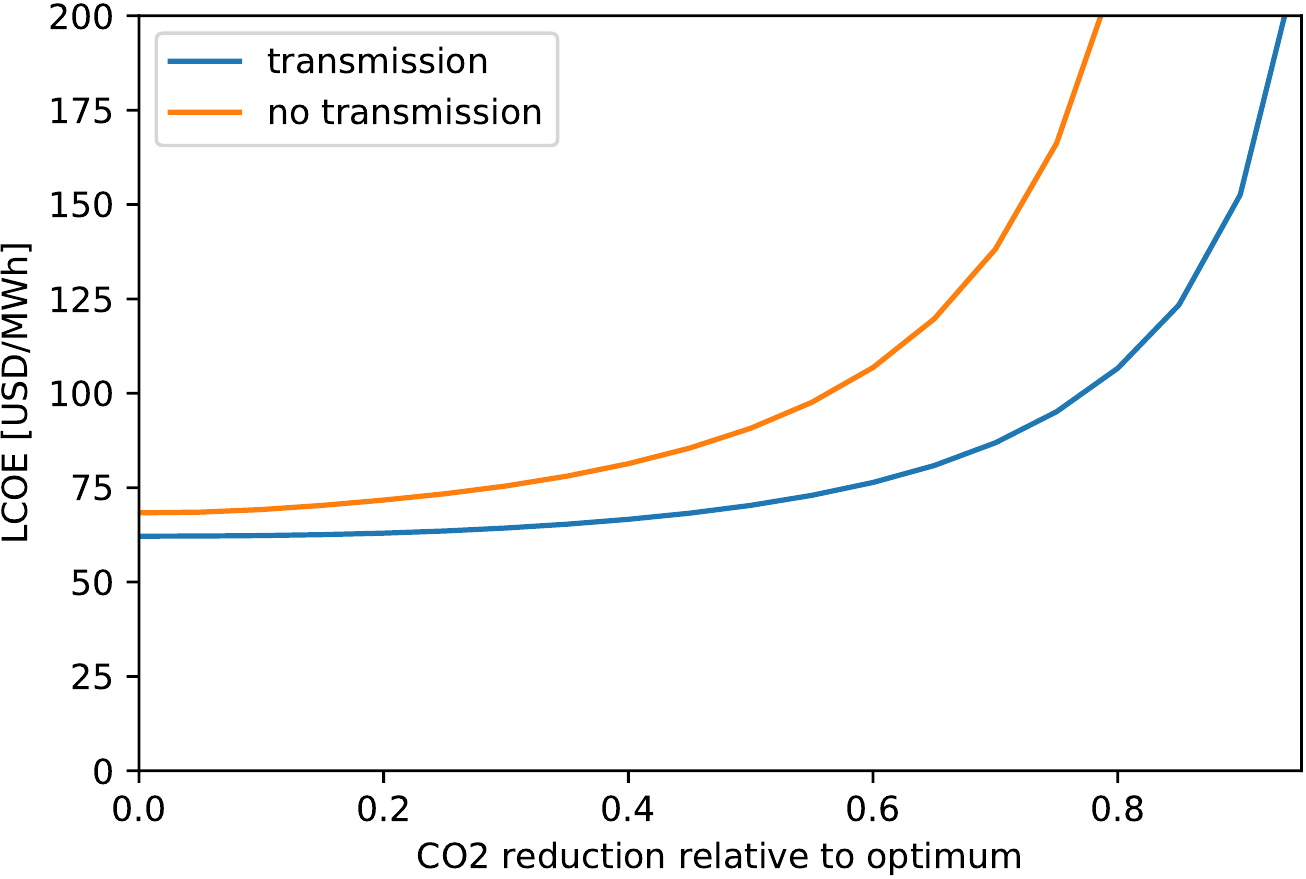}}
\caption{\label{fig:prices}Levelised cost of electricity in dependency of the $CO_2$ reduction relative to the cost-optimum without any constraints with and without transmission grid.}
\end{figure}

The optimised generation mix for both transmission scenarios is depicted in Fig. \ref{fig:distrb}. 
With transmission, the major share of wind generation facilities is installed in the South and partly the central North, whereas solar PV is installed in central Vietnam.
This comes jointly with a strongly reinforced transmission grid in southern Vietnam, while the connection to the northernmost region is comparably weak.
In addition, no renewable resources are installed in the uppermost north, where capacities are less cost-effective.
It should be noted that OCGT plants are not effected by location-dependent availability. Consequently, their production capabilities are independent of their location (with transmission grid restrictions) and therefore
the optimum is flat with respect to the spatial distribution of OCGT capacities. \\
Without transmission, the picture changes partially. All nodes have similar shares of solar PV of 25\% to 35\%, while the remaining generation is provided mostly (for some nodes entirely) by OCGT.
Nodes with comparably large wind shares are still allocated these, thus indicating the resource potentials being responsible for the installations instead of the topological position within the network.
This is also suggested by the assumption that transmission costs are almost entirely neglected, hence rendering transmission grid expansion cheap.\\
To increase the shares of renewables in the system, the CO$_2$ emission constraint (Eq. \ref{eq:co2}) is introduced and tightened.
The resulting generation mixes in dependency of the CO$_2$ reduction relative to the cost-optimal case without constraints are shown in Fig. \ref{fig:optmix}
for both transmission scenarios.
For OCGT and solar PV, the picture looks similar in both cases.
Shares of solar PV remain, especially with transmission, roughly constant, while OCGT is gradually replaced by wind.
However, the cost-optimal share of wind generation is at 38\% much larger with the fully optimised transmission grid than in the case without transmission (11\%).
This represents the fact that wind is less correlated on the spatial scale than PV and therefore benefits much stronger from a transmission grid (\cite{kies2016curtailment}).
\\Fig. \ref{fig:prices} shows levelised cost of electricity (LCOE) in dependency of the CO$_2$ reduction relative to the optimum distribution without any constraints for both transmission grid scenarios.
With optimised transmission, LCOE is at 62 USD/MWh considerably below marginal cost of dispatchable generation, thus indicating the high favourability of renewable generation as well as a significant importance of transmission for
the Vietnamese power system.
Without transmission, LCOE is at 68 USD/MWh in the optimal case, but costs rise drastically, if the CO$_2$ constraint is tightened.

\section{Critical appraisal}
A number of simplifications made might effect the conclusions presented in this paper.\\
In the paper, we have not considered hydro power as part of the Vietnamese power supply.
However, hydro power is considered fairly well exploited in Vietnam with installed hydro power capacities expected to grow (according to the Vietnamese power development plan) from 13.6 GW in 2014 to 20.8 GW in 2030.
This compares to an overall growth of installed capacities from 34 GW to 116 GW.
Hence, hydro power might have the potential to complement the renewable mix in an extraordinary beneficial way and it is planned to consider hydro power using an potential energy approach to calculate energy inflow into hydro power storages in future versions of this work.
\\Neither hydro nor any other possibility to store energy over time is incorporated in the presented work.
Hence, in future work medium-scale storage technologies like batteries will be investigated to quantify their potential impact on the future Vietnamese power system.
However, such storage solutions are usually not cost-competitive today and it is difficult to model their future characteristics (cost, efficiencies, degradation in case of battery storage, etc.).
In addition, degradation processes of batteries are highly non-linear.
\\We have also only considered a single year (meteorological year 2012) so far. Therefore, results might only represent this year and not be generalizable to a longer temporal period.
\\It is planned to address all of the aforementioned deficiencies in future work.
\section{Summary, conclusions and outlook}

This paper investigates the integration of large shares of fluctuating generation from the renewable sources of wind and photovoltaics into the Vietnamese power system.
\\It is shown that wind and solar PV can provide more than two thirds of the overall generation, if the transmission grid is sufficiently strong. 
However, for even higher shares the possibility to shift generation in time, i.e., storage, is required.
In many countries around the world, significant energy storage is provided by 
hydro power and Vietnam already uses a fair share of hydro power.
Therefore, a straightforward extension of this work is to include hydro power into calculations and to investigate the question, to what
extent existing hydro power facilities can contribute to the system integration of renewables in Vietnam.
\\The transmission grid has the potential to reduce LCOE in a highly renewable Vietnamese power system by around 10\% caused by the large benefit for wind power that is provided
by the transmission network. The advantage of wind over PV from transmission can be concluded from the large increase of wind shares in the optimal solution with vs. without transmission.
To distangle the optimum caused by resources and by network topology, more investigations towards the cost sensitivity in case of transmission grid expansion are required.
\\Together with the fact that renewable generation capacities were mostly installed in southern Vietnam and that major demand centres are in southern and northern Vietnam, the results emphasize the potentials of renewables in Vietnam together with an expanded transmission grid.
This cost-optimal system has LCOE of approximately 62 USD/MWh. However, investment and operational costs for the transmission grid are not included in this number.
\\This work has highlighted the potentials of fluctuating renewable generation from the sources of wind and solar PV in a future Vietnamese power system.
In a forthcoming study, the methodology and database will be expanded to cover more meteorological years to reduce the sensitivity towards potential extreme events in the data, 
and to include more technologies such as hydro power, energy storage, etc. 
In addition, it is planned increase the spatial detail and also consider existing official expansion plans of Vietnam's electricity grid.
\section*{Acknowledgements}

This work is part of the R\&D Project “Analysis of the Large Scale Integration of Renewable Power into the Future Vietnamese Power System” financed by Gesellschaft fuer Internationale Zusammenarbeit GmbH (GIZ, 2016-2018).
Furthermore, A. Kies is financially supported by Stiftung Polytechnische Gesellschaft.
A slightly different version of this paper has been published in Energy Procedia (\cite{kies2017large}).


\bibliography{xampl2}
\bibliographystyle{unsrt}
\end{document}